\documentclass[aps,prl,showpacs,floatfix,twocolumn]{revtex4}
\usepackage{amsmath}
\usepackage{amsfonts}
\usepackage{amssymb}
\usepackage{color}
\usepackage{graphicx}
\usepackage{subfigure}
\usepackage{verbatim}

\newcommand{\romd}{{\text{d}}}

\begin{document}

\title{Morphogenesis of membrane invaginations in spherical confinement}
\author{Osman Kahraman$^{1}$, Norbert Stoop$^{2}$, and Martin Michael M\"{u}ller$^{1}$}
\affiliation{$^{1}$Equipe BioPhysStat, ICPMB-FR CNRS 2843, Universit\'{e} de Lorraine; %
1, boulevard Arago, 57070 Metz, France}%
\affiliation{$^{2}$Computational Physics for Engineering Materials, ETH Zurich, %
 Schafmattstr.\ 6, HIF, CH-8093 Zurich, Switzerland}%
\date{\today}

\begin{abstract}
We study the morphology of a fluid membrane in spherical confinement. When the area of the membrane is 
slightly larger than the area of the outer container, a single axisymmetric invagination is observed. For higher 
area, self-contact occurs: the invagination breaks symmetry and deforms into an ellipsoid-like shape connected to 
its outer part via a small slit. For even higher areas, a second invagination forms inside the original invagination. The folding 
patterns observed could constitute basic building blocks in the morphogenesis of biological tissues and organelles.
\end{abstract}


\maketitle


\paragraph{Introduction.}
Folding phenomena are ubiquitous in nature and living matter. 
They are a key to understanding the complex shapes of the mammal gut \cite{Savin2011}, 
the cerebellum \cite{Sudarov2007} and the kidney \cite{Kuure2000}, and are of fundamental importance in 
developmental biology.
 
Typically, folding processes are triggered by either a reorganization of mass due to 
growth \cite{Dervaux2009,Mahadevan2009,Hannezo2011} 
or by a buckling instability originating from external forces or constraints \cite{BenPom1997,Gompper2006,Witten2007}. 
Here, we study the folding deformations of a thin closed membrane inside a spherical cavity. 
This system can be considered a prototype for surface invagination, a process that occurs in different 
biological systems. 
Among others, well-known examples are the gastrulation within the surrounding egg 
shell \cite{Labouesse2011} or the crista formation of the inner mitochondrial membrane inside the outer 
membrane \cite{mannella2006}. In both cases, the fold formation is a geometric necessity, since the inner 
surface is too large to fit into the cavity. Scientists have studied the resulting shapes and how they develop during growth 
since decades.

In this Letter, we use a minimal mechanical model of a fluid membrane to study the shape of such invaginations. 
Using finite element simulations, we construct a morphological phase diagram showing what shapes 
emerge for given membrane area and enclosed volume. We find an initial invagination of axisymmetric 
type for small surface growth, which breaks symmetry into ellipsoidal shapes as the membrane surface is increased. 
For even higher prescribed area, a secondary invagination emerges inside the first one as a result of the 
interplay of volume and surface constraints together with self-contact. Our analysis is capable of showing the 
morphogenesis of cristae-like invaginations with as few ingredients as possible, and extends previous 
works of, e.g., Ref. \cite{renken2002} considerably.


\paragraph{The model.}
The classical curvature model developed by Canham, Helfrich, and Evans expresses the mechanical energy
of a membrane related to its bending as a second order expansion in curvatures \cite{Canham,Helfrich,Evans}:
\begin{equation}
E_b = \frac{\kappa}{2}\int \romd A \, (2H-C_0)^2 + \kappa_G \int \romd A \, K 
\; ,
\label{eq:bendingenergy}
\end{equation}
where the integrals are carried out over the surface of the membrane. Here, $H$ is the mean curvature and $K$
is the Gaussian curvature. The spontaneous curvature $C_0$ represents an intrinsic preferred mean curvature 
of the membrane which we will set to zero for simplicity.
$\kappa$ and $\kappa_G$ denote bending and Gaussian rigidity, respectively.
For a given topology, the second term in the energy~(\ref{eq:bendingenergy}) equals a constant due to the Gauss-Bonnet 
theorem. We can omit this term here since only closed membranes with spherical topology are considered.
Fixing area and volume to $\bar{A}$ and $\bar{V}$, respectively, the scaled total energy of the membrane can thus be written as
\begin{equation}
\tilde{E} = \int \romd A\, 2 H^2 + \frac{\tilde{\mu}_A}{2}(A-\bar{A})^2 +  \frac{\tilde{\mu}_V}{2}(V-\bar{V})^2
\; ,
\label{eq:helfrich}
\end{equation}
where $\tilde{\mu}_A$ and $\tilde{\mu}_V$ are penalty factors implementing the constraints. 
This model is not only relevant for fluid lipid bilayer membranes but can also be applied to growing soft tissue 
as long as in-plane shear forces can equilibrate on time scales smaller than growth \cite{Gorielyetal2008}.

We mimic the effect of the external constraint by adding a soft spherical container to the system.
This container is modeled via a spherical force field which penalizes the motion of the membrane
outside the container. 
Applying such a soft constraint implies that the membrane is allowed to trespass into the force 
field at the expense of increasing its energy. As a result, the radius of the overall shape can be slightly 
larger than the radius of the outer shell during the simulations.


\paragraph{Finite element simulations.}
The bending energy involves squares of curvatures, which are second derivatives of the surface vector function. 
For a finite element simulation it is possible to use either a mixed method as suggested in Ref.~\cite{stinner2010} or 
trial functions that have square integrable derivatives up to second order. An elegant formulation of such elements 
for thin shells has recently been developed by Cirak et al. The method is based on a CAD subdivison 
scheme which satisfies this $C^1$ continuity requirement of the surface vector function \cite{cirak2000,cirak2001}. 
Klug and coworkers extended this approach to fluid lipid membranes \cite{feng2006,ma2008}. 
Following these works we discretize the scaled membrane energy given in eq.~(\ref{eq:helfrich}) 
and set up corresponding forces that act on the nodes of the mesh. To prevent mesh distortions and possible numerical 
instabilities, we applied viscous regularization and r-adaptive remeshing schemes as described in \cite{ma2008}. Instead 
of using line search based solvers as in Ref.~\cite{ma2008},
we performed a time integration and added a damping force on the nodes to equilibrate the system analogous to 
the method used in Ref.~\cite{stoop2010} for thin shells.


\paragraph{Moderate surface growth}
We simulated the membrane in a spherical container of unit size for different values of scaled surface area 
$a=\bar{A}/A_0$ and scaled volume $v=\bar{V}/V_0$, where $A_0$ and $V_0$ are the area and the volume of 
the container. 

\begin{figure}
\begin{center}
  \subfigure[][]{\label{fig:equilibriumsolution}\raisebox{0.3cm}{\includegraphics[width=0.23\textwidth]{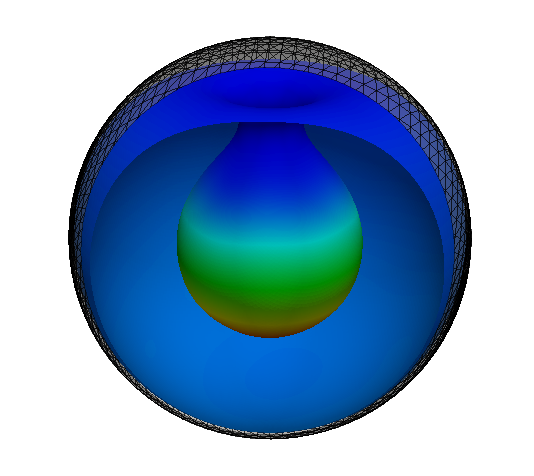}}}
  \subfigure[][]{\label{fig:slice}\includegraphics[width=0.23\textwidth]{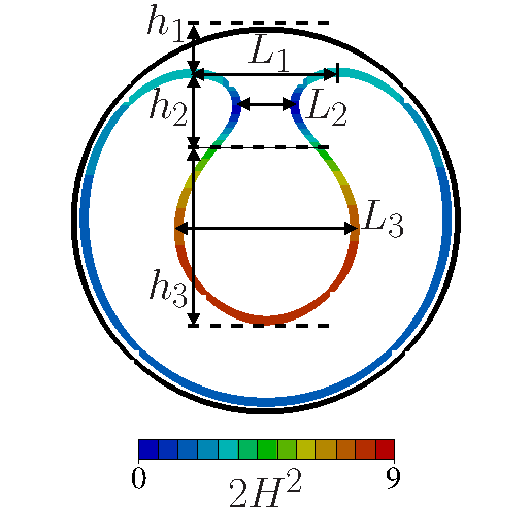}}
\end{center}
\caption{(a) Numerical equilibrium solution for a membrane (smooth surface) inside a spherical container (black 
mesh) with scaled area $a=1.2$ and scaled volume $v=0.8$. 
The membrane bulges inward and forms an invagination reminiscent of a light bulb. 
(b) Vertical slice of the system, which contains the axis of symmetry, together with the corresponding bending energy density 
of the membrane ($\equiv 2H^2$). 
In (b) the measured distances of the invagination are defined (see Tab.~\ref{tab:measurements}). 
\label{fig:system}}
\end{figure}
If the area of the membrane is larger than the area of the container, the membrane has to fold into the interior. 
For moderate values of surface growth, the membrane forms a single invagination, connected via a neck to the 
part in contact with the outer shell (see Fig.~\ref{fig:system}).  Every additional invagination 
would contribute an extra energy of about $8\pi\kappa$---the value of the bending energy of a sphere---to the overall 
bending energy which is why multiple invaginations are not observed in equilibrium. 

The shapes that we obtain for moderate surface growth appear to be axisymmetric. To confirm this, we took horizontal slices
of the surface and analyzed their curvature: first, we determined the symmetry axis by taking the mean of the surface 
normals of each vertex. Second, we took a slice of the membrane perpendicular to the 
axis of symmetry at the point where the invagination is the thickest. 
Using a discrete formulation of the curvature based on a second order polynomial approximation of the curve on sample 
points \cite{lewiner2005}, we 
estimated the curvature for each invagination slice. The results confirm that the obtained shapes are axisymmetric for moderate 
surface growth. In this case it is thus sufficient to consider a two-dimensional vertical 
slice of the shape which contains the axis of symmetry. The corresponding slice for $(a,v)=(1.2,0.8)$  
is shown in Fig.~\ref{fig:slice} together with the bending energy density. 
Although the membrane shape resembles a stomatocyte as is found in the corresponding reduced volume 
problem without container, we note that it is the confinement which forces the membrane 
into this form; a free vesicle would adopt the form of an ellipsoid for these parameter values \cite{Seifert}. 

\begin{figure}
  \centering
  \includegraphics[width=0.48\textwidth]{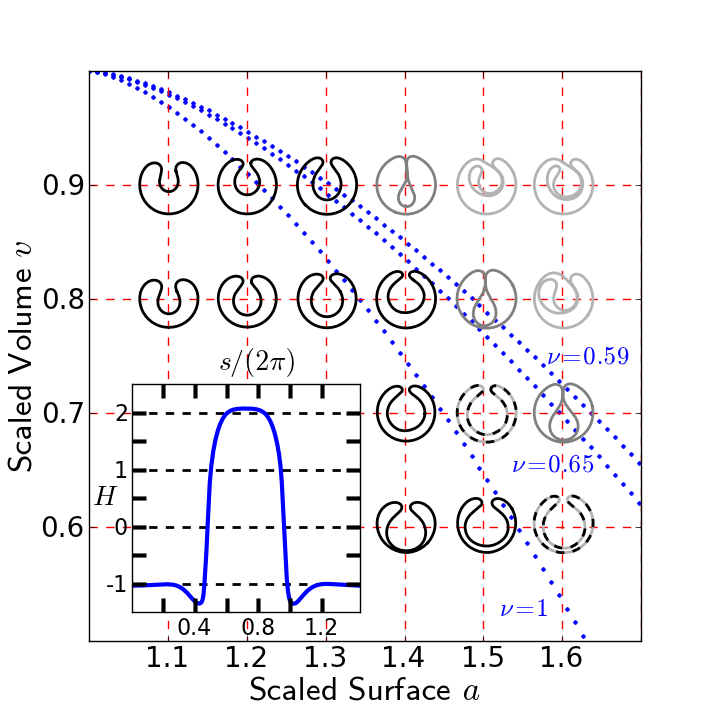}
  \caption{Morphological phase diagram with vertical slices of each simulation point. The color of each slice corresponds to the different regimes observed: 
  the invagination is either axisymmetric (black), ellipsoid-like (dark gray), or stomatocyte-like (light grey). The slices of the non-axisymmetric shapes 
  lie in the symmetry plane perpendicular to the slit-like neck. 
For the dotted slices a metastable ellipsoid-like state has been observed. 
Eq.~(\ref{eq:theoretical estimate}) is plotted for different values of $\nu$ (dotted lines).
  \textit{Inset:} mean curvature $H$ of the vertical slice $(a,v)=(1.2,0.8)$ as a function of scaled arc-length $s/(2\pi)$.}
  \label{fig:phasediagram}
\end{figure}

\begin{table}
     \centering
\begin{tabular}{|c||c|c|c|c|c|c|c|}
  \hline
$(a,v)$ & $L_1$ & $L_2$  & $L_3$ & $h_1$ & $h_2$ & $h_3$ & $e_b$
\\ \hline 
(1.1,0.9) &  0.97  & 0.48  & 0.65  & 0.25 & 0.40 &  0.59 & 1.91
\\ \hline 
(1.2,0.9)  &  0.69  & 0.26  & 0.86  & 0.13 & 0.35 &  0.87 & 1.97
\\ \hline 
(1.3,0.9) &  0.65  & 0.27  & 0.97  & 0.07 & 0.37 &  1.07 & 2.00         
\\ \hline \hline
(1.1,0.8)  &  1.00  & 0.48  & 0.76  & 0.36 & 0.44 &  0.72 & 1.95
\\ \hline 
(1.2,0.8)  &  0.76  & 0.28  & 0.95  & 0.23 & 0.36 &  0.97 & 1.98
\\ \hline   
(1.3,0.8)  &  0.69  & 0.24  & 1.08  & 0.15 & 0.35 &  1.11 & 1.99
\\ \hline 
\end{tabular}
\caption{Measurements of characteristic system parameters [see Fig.~\ref{fig:slice}]. 
The error of measurement is about $\pm 0.05$ for all parameters including the scaled bending energy 
$e_b:=E_b/(8\pi \kappa)$.}
\label{tab:measurements}
\end{table}
To analyze the shapes quantitatively, we examine the corresponding system parameters defined in Fig.~\ref{fig:slice}: 
the variable $L_1$ denotes the horizontal distance between the uppermost points, and $L_2$ and $L_3$ are taken at 
the narrowest and the broadest point of the invagination, respectively. The vertical distance $h_1$ 
indicates how much the membrane detaches from the container; $h_2$ measures the length of the neck between the uppermost points and 
the points where the curvature of the slice changes its sign. The distance $h_3$ is the extension of the tip of the invagination. 
In Tab.~\ref{tab:measurements} the corresponding numerical values are listed for $v=0.8$ and $0.9$. We remark 
that for constant volume $v$ and increasing surface area $a$, the length $h_3$ of the tip increases, whereas the neck decreases in length $h_2$ and diameter $L_2$. 
Increasing the inner volume from 0.8 to 0.9 does not significantly change the overall shape, but causes the invaginations to penetrate less 
(see Fig.~\ref{fig:phasediagram}).

We have also measured the bending energy $e_b$ of the whole membrane shape, normalized by the bending energy 
of a sphere (see again Tab.~\ref{tab:measurements}). We find that $e_b$  is very close to two, $i.e.$, the energy of a system of 
two isolated spheres.  
We therefore consider such a system as a simple model for the membrane. In this case, there is \textit{no} connection 
between the invagination ($i.e.$, the inner sphere) and the part of the membrane in contact with the container 
(the outer sphere). The radius of the outer sphere is one, the radius of the inner sphere is set to $R_i$. The ratio $a$ of total area to the area 
of the container can be written as $a = 1 + R_i^2$. 
Similarly, the ratio $v$ of the volume between the two spheres to the volume of the container is given by
$v = 1 - R_i^3$. 
Hence, we get the following simple relation between the two parameters:
$v = 1 - (a-1)^{3/2}$.
If this equation is fullfiled, both constraints can be accomodated by the 2-sphere system. 
At $a=1$ and $v=1$, the inner radius vanishes and we are just left with
one unit sphere. At $a=2$ and $v=0$, the inner sphere reaches
the size of the outer one, and the system consists of two unit spheres.
If $v<1 - (a-1)^{3/2}$, the system will be stressed, since the volume 
between the two spheres wants to be smaller than the membrane can 
accomodate with the available area. Na{\"i}vely, one would think that no axisymmetric ground state exists any more. 
In the simulations, however, the outer part can detach from the container. This allows 
the membrane to decrease the enclosed volume such that an axisymmetric state is still energetically favored. 

A visual inspection of the shapes obtained suggests that the neck can be approximated theoretically by a 
catenoidal batch connecting the two spheres. However, such a model turns out to be 
insufficient to explain the results of the simulations \textit{quantitatively}. 
A quick glance at the mean curvature along a vertical simulation slice explains why [see inset of 
Fig.~\ref{fig:phasediagram}]: the mean curvature $H$ at the neck is not close to zero but changes abruptly from negative 
(part in contact with the container) to positive (invagination).

\begin{figure}
  \centering
  \subfigure[][]{\label{fig:prolate}\includegraphics[width=0.22\textwidth]{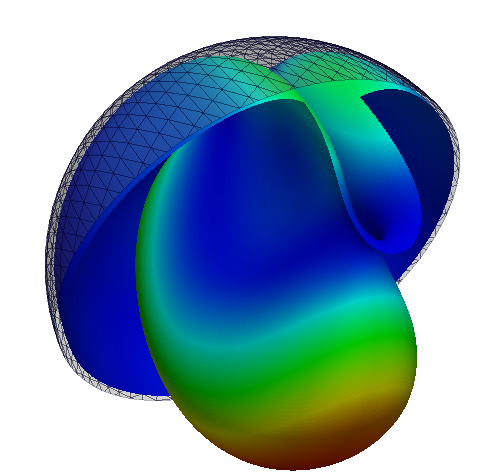}}
  \subfigure[][]{\label{fig:stoma}\includegraphics[width=0.22\textwidth]{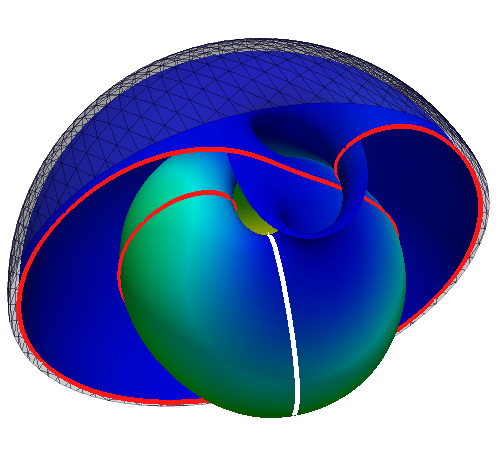}}
  \\
  \subfigure[][]{\label{fig:stomacut1}\includegraphics[width=0.22\textwidth]{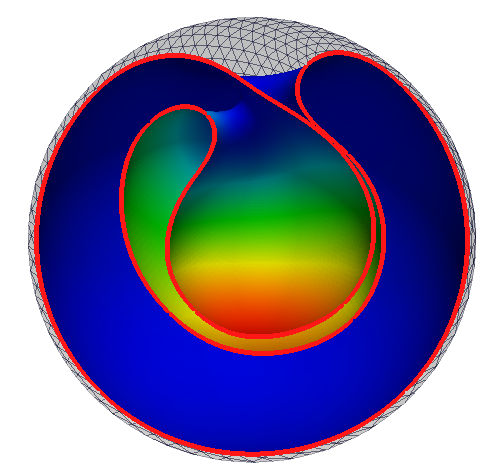}}
  \subfigure[][]{\label{fig:stomacut2}\includegraphics[width=0.22\textwidth]{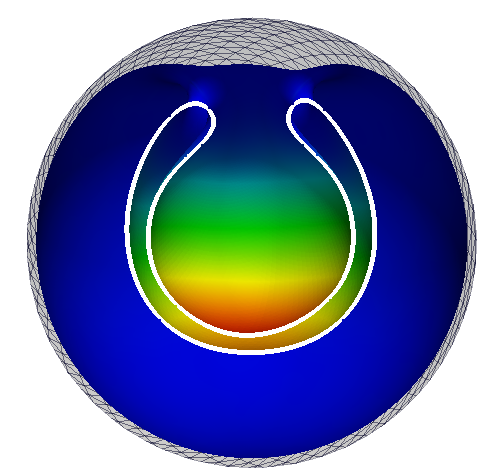}}
  \caption{Numerical equilibrium solutions which break axisymmetry. (a) Ellipsoid-like state for $(a,v)=(1.6,0.7)$. 
(b) Stomatocyte-like state for $(a,v)=(1.6,0.9)$. (c) Cut of the stomatocyte-like state along the symmetry plane. 
(d) Cut of the stomatocyte-like invagination perpendicular to the symmetry plane.}
  \label{fig:highsurface}
\end{figure}

For $v>1 - (a-1)^{3/2}$, too much volume wants to be stuffed into a membrane of limited area. 
One may anticipate that this frustration is resolved by breaking symmetry. This happens indeed as one can see in Fig.~\ref{fig:prolate}.
However, the neck mitigates this effect: the tip of the invagination can still keep a shape close to a sphere if $v$ is not too 
large (see Fig.~\ref{fig:phasediagram}).


\paragraph{High surface growth and symmetry breaking}
Packing increasing membrane area into a constrained volume inevitably produces self-intersections. 
To avoid this, we implemented a collision handling algorithm based on the minimization
of the intersection contour of polygonal elements \footnote{
Once a polygonal element of the mesh intersects with another one, 
a linear gradient vector is calculated for each of its penetrating edges \cite{volino2006}. 
This gradient vector points in the direction along which the 
corresponding edge should be displaced in order to minimize the intersection
contour of the two polygons. A scaling of this vector yields a force vector
that we distribute over the nodes of the penetrating polygon. Hence, every self-contact of 
the mesh is detected as soon as it occurs and penalized by forces that untangle it smoothly.}.

With this algorithm, non-axisymmetric systems can be studied as well. For $(a,v)=(1.5,0.7)$, and 
$(1.6,0.6)$---$i.e.$, systems with an axisymmetric global minimum---a metastable 
symmetry breaking state with self-contacts has been observed in the simulations.
At the connecting neck, the membrane touches itself and forms a non-spherical slit, which endures for long simulation 
times, indicating that this configuration is a local minimum.
The invagination is not spherical in this case. Its form is reminiscent of the oblate and prolate shapes found 
in reduced volume problems such as in \cite{Seifert}.

To include this behaviour in the theoretical model, we again neglect the neck and assume that the system consists of 
an outer sphere of radius one ($i.e.$, the part of the membrane in contact with the container) and a vesicle of volume $\nu 4\pi R^3_i /3$
(the invagination), where the parameter $\nu \in \{0,1\}$ is the reduced volume which measures the deviation from the volume
of a sphere. The relation between $a$ and $v$ now reads:
\begin{equation}
  v=1-\nu(a -1)^{3/2}
  \label{eq:theoretical estimate}
  \; .
\end{equation}
According to Ref.~\cite{Seifert} the theoretical model adopts the following shapes:
for $0.65\le\nu\le 1$ the global energy minimum corresponds to a spherical volume enclosing a prolate ellipsoid in the middle.  
Decreasing the value of $\nu$ more and more, causes the ellipsoid to become oblate for $0.59\le\nu\le 0.65$,
while for $\nu\le 0.59$ the invagination distorts into a stomatocyte shape.
In our simulations corresponding shape transitions can indeed be found  (see dotted lines in Fig.~\ref{fig:phasediagram}). 
The connection via the neck, however, alleviates these transitions. Moreover, the confinement gives rise to more complex invagination geometries: 
we distinguish two different types of non-axisymmetric invaginations in the simulations. 
Ellipsoidal invaginations have a complex slit-like neck: the middle of the slit touches itself so that it appears as if there 
are only two holes as openings [see Fig.~\ref{fig:prolate}]. 
Increasing the surface area of the membrane or the volume
enclosed by it decreases $\nu$; a second invagination bulges into the first one [see Fig.~\ref{fig:stoma}--\ref{fig:stomacut2}].
The shape of this secondary invagination resembles a stomatocyte [see Fig.~\ref{fig:stomacut2}, white line] connected to 
the outer part via a slit-like neck.


\paragraph{Conclusions}
Combining a simple mechanical model of a closed fluid membrane with the constraint
of a confining shell, we investigated the morphogenesis of membrane invaginations.
A theoretical approximative model allowed us to classify the shapes in a morphological phase diagram. 
Owing to the constraint, the equilibrium shapes deviate considerably from the reduced volume solutions 
found for membrane vesicles without confinement \cite{Seifert}. 
The axisymmetric invaginations obtained for moderate surface growth 
resemble closely the basic geometry of biological invaginations such as
inner mitochondrial cristae or embryonal gastrula. 
Incorporating the treatment of self-contacts of the membrane into the model, we found non-axisymmetric 
shapes for high surface growth, namely ellipsoid-like and stomatocyte-like invaginations. 
Our computational analysis of a minimal membrane model could help to establish which structures 
are intrinsic to a folded membrane and which may be particular to other effects such 
as membrane protein interactions in the case of mitochondria \cite{voeltz2007, rabl2009}
or localized cell differentiation in gastrulation \cite{Lecuit2007}.


\paragraph{Acknowledgements}
Financial support from the University of Metz is acknowledged. The authors would like to thank Jemal Guven, 
Herv\'e Mohrbach, Gaetano Napoli, and Troy Shinbrot for helpful discussions.



\end{document}